\documentclass{conm-p-l}
\newtheorem{theorem}{Theorem}[section]
\usepackage{graphicx}
\title{Discrete spacetime and its applications}
\author{Eitan Bachmat}
\newcommand\Var{\ensuremath\operatorname{Var}}
\newcommand\Vol{\ensuremath\operatorname{Vol}}
\begin{document}
\maketitle
A discrete spacetime is a random partially ordered set. It is obtained by sampling $n$ points from 
a compact domain in a spacetime manifold w.r.t.\ the volume form. The partial order is given by restricting the causal 
structure on the domain to the sampled points. These objects were introduced by physicists in an attempt 
to reconcile gravity with quantum mechanics, \cite{Myrheim,Hooft,BLMS}. They received very little direct 
attention in the mathematics literature.

Recently, discrete spacetimes have appeared in very diverse applications which are
very far removed from their original motivation.
In this self-contained survey we will present some results and numerical calculations regarding the asymptotic behavior of these objects. 
Some of the more basic results are stated as theorems since they are particularly useful in the applications. 
Numerical calculations are presented with the hope that they will incite some theoretical work, 
aimed at proving some related assertions.
We then present a few of the applications by shamelessly following the pattern of the 
recent survey paper \cite{De}, which describes some related material. We will present a short list of seemingly unrelated problems 
and show how they fit into the discrete spacetime formalism. 
While this is a survey paper, it does contain some new material, namely, the statement of theorem~\ref{antichain}, the (simple) 
solution to the dimension reconstruction problem, the numerical calculations of figure~\ref{fig:TW}, 
the connect-the-dots application and the relation between discrete spacetime 
and the literature on maximal layers. 
A more comprehensive and extensive treatment of the subject along with full proofs of the basic 
results will be presented in a forthcoming publication. 

Our own research relating to discrete spacetime owes much to the constant encouragement which was provided by Percy Deift
and it is our sincere pleasure to dedicate this paper to him on the occasion of his 60th birthday. 

\section{The asymptotic behavior of discrete spacetime}
A spacetime (Lorentzian manifold) $M$ is an $m=d+1$ dimensional real, 
connected, $C^{\infty}$ manifold with a globally defined $C^{\infty}$
symmetric tensor field $g$, of type $(0,2)$, which is non-degenerate and Lorentzian. By Lorentzian 
we mean that for any $a\in M$ there is a basis of $T_a(M)$, the tangent space to $M$ at 
$a$ relative to which $g_a$ is given by the diagonal matrix with diagonal entries $(-1,1,1,\ldots ,1)$. 
We consider a set of local coordinates $x_0,\ldots , x_d$ on the manifold. In terms of the local coordinates it is 
convenient to consider the expression
\begin{equation}
\label{metric}
ds^2=\sum_i\sum_jg_{i,j} \, dx_i \, dx_j
\end{equation}
The metric defines a canonical volume form which, using the coordinates, may be written as 
\begin{equation}
\label{volume}
\Vol_a=\sqrt{\vert \det(g_a)\vert }\,d^mx
\end{equation}
The metric $g_a$ also classifies the tangent vectors $v\in T_aM$ into {\it timelike} vectors, if $g_a(v,v)>0$, {\it null} or 
{\it lightlike}, if $g_a(v,v)=0$
and {\it spacelike} if $g_a(v,v)<0$. The set of timelike vectors consists of a double cone. We say that a spacetime is time orientable if there 
is a non-vanishing vector field $V$ consisting of timelike vectors. We will assume that all our spacetime manifolds are time orientable. An orientation provides a continuous choice of one out of the
two cones (the one containing the vector $V_a$) and we declare this cone to be the future-pointing cone. A parametrized curve in the manifold
is said to be {\it future-pointing timelike} if all its tangent vectors belong to the future-pointing cone.

The simplest spacetime manifold is flat Minkowski space, which is ${\mathbf R}^m$ equipped with the constant metric
\begin{equation}
\label{minkowski}
ds^2=dx_0^2-dx_2^2-dx_3^2-\ldots -dx_d^2
\end{equation} 
Other natural examples to consider are the constant curvature  hyperbolic (anti de-Sitter) and spherical (de-Sitter) spacetime manifolds. In two dimensions, hyperbolic spacetime with constant curvature $-1$ is given by the metric
\begin{equation}
\label{hyper}
ds^2=\frac{1}{cos^2(y)}(dx^2-dy^2)
\end{equation}
where $-\pi /2<y<\pi /2$. Spherical spacetime is realized on the same manifold by simply inverting the sign in (\ref{hyper}).

Consider a compact domain $D\subset M$. For the applications we have in mind it is sufficient to consider very ``nice'' domains,
without global pathologies, or intricate smoothness issues, 
so we will assume for simplicity that the interior of $D$ is an $m$ dimensional, contractible, submanifold
of $M$ and that the boundary of $D$ is $C^{\infty }$, except possibly 
at a lower dimensional $C^{\infty }$ submanifold.  
The causal relation (past-future) is defined on pairs $a,b\in D$ by $a\prec b$ iff there is a future-pointing timelike curve in $D$ from $a$ to $b$.
We will also assume that the relation $\prec $ is a partial order relation, namely, there are no points 
$a\neq b\in D$ with future-pointing timelike curves (in $D$) from $a$ to $b$ and from $b$ to $a$. 

Given a spacetime 
$M$ and a pair of points $a\prec b$, a natural domain to consider is the interval $[a,b]$ 
consisting, of all points $z$ which satisfy 
$a\prec z\prec b$. 
\begin{figure}[tb]
\centerline{
\includegraphics{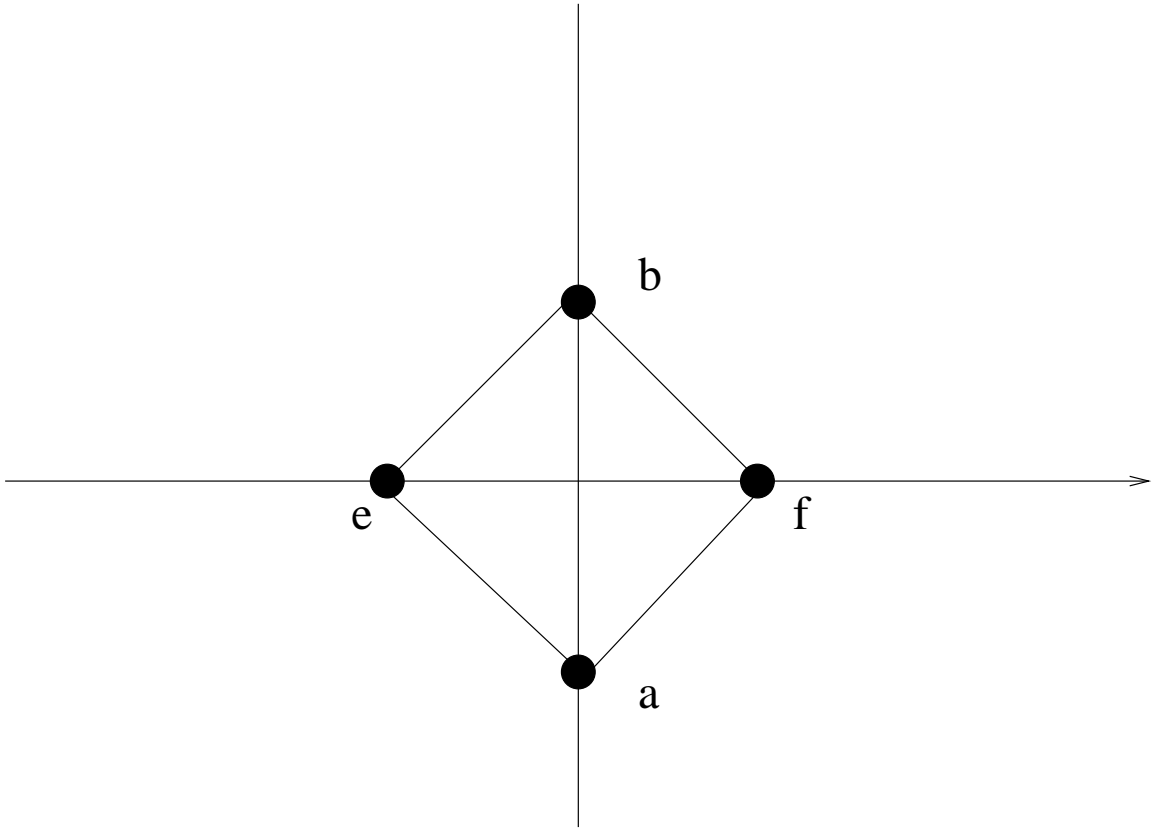}
}
\caption{An interval $[a,b]$ in the Minkowski plane.}
\label{fig:rhombus} 
\end{figure}
The interval $[a,b]$ for a pair of points in the Minkowski plane is shown in figure~\ref{fig:rhombus}, as the tilted square 
$\vert x+y\vert \leq \delta$, whose vertices are $a,b,e,f$. The same domain doubles as the interval of $[a,b]$ in spherical spacetime
if $\delta <\pi /2$ and of $[e,f]$ in hyperbolic spacetime (restricting to $\vert y\vert <\pi /2$). The interval $[e,f]$ is a compact domain in hyperbolic spacetime if $\delta <\pi /2$.
 
A {\it discrete spacetime} associated with a compact domain $D$ is a random partially ordered set on $n$ points obtained by sampling the $n$ points from $D$ using the volume form and inducing the relation $\prec $ on the chosen points. 

We are interested in the asymptotic properties of discrete spacetimes associated with a domain 
$D$ as $n$ becomes large. In particular,
we may think of a discrete spacetime as a finite approximation to $D$. It is then natural to ask 
whether we can ``reconstruct'' $D$
from larger and larger discrete spacetimes associated with it. 

At first this seems difficult since we have thrown away the original metric and retained only the causal 
relation and the volume form. However, results of Hawking, King and McCarthy, \cite{HKM}, and of Malament, \cite{Ma} 
show that, in the continuous setting, knowledge of causality and the volume form is essentially equivalent to knowledge of the metric.

We can consider more closely the type of quantities that we may want to reconstruct.
Given a timelike curve $C$ we may define its length $L(C)$, better known as {\it proper time}, to be $\int_C ds$. 
In relativity theory, this parameterization independent quantity 
measures the time which passes on a clock which is attached to a point-like object whose trajectory is given by the curve $C$. Proper time is also 
the action for the trajectory of a point-like object between two points $a,b\in M$. This means that a 
free falling point-like object moving between events $a$ and $b$ will follow a critical path for this action, this is Einstein's 
equation of motion. The critical paths, which
are maximizing in this case, are the timelike geodesics in analogy with the situation in Riemannian geometry 
(where they are minimizing). 

Assume that $a,b\in D$, with $a\prec b$. We may define their ``distance'' in $D$ to be $L_D(a,b)=\sup_C \; L(C)$, where the supremum 
is taken over all timelike curves $c\subset D$, whose initial point is $a$ and whose endpoint is $b$. 
It is known that $L_D(a,b)$ is finite when $D$ is compact and is
achieved by a timelike curve, \cite{Pe}. We shall call such a curve a {\it maximal curve}. A maximal curve may have non-geodesic 
segments on the 
boundary of $D$,
but in the interior of $D$ it must satisfy the geodesic equation. It must also be differentiable at all points where the 
boundary is differentiable. 

We would like to have discrete analogues of the concepts of length and maximal curves.
Given a partial order $\prec $, a {\it chain } of size $k$ is a set of elements $x_1,\ldots ,x_k$ such that $x_i\prec x_{i+1}$ for all
$1\leq i<k$. A chain in a discrete spacetime $P$ associated with $D$ is the obvious discrete analogue of a timelike curve. 
Under this correspondence the analogue of curve length is simply the size of the chain, and the analogue of a maximal curve between 
$a$ and $b$ will be a maximal sized chain $x_1,\ldots ,x_k$ in $P$ such that $a\prec x_1$ and $x_k\prec b$. 
We let $L_P(a,b)$ denote the length of such a maximal sized chain.    

We will say that an event occurs with high probability (w.h.p.) if it occurs with probability approaching~1
as $n$ tends towards infinity.
The following folk theorem provides the asymptotic high probability link between the discrete distance $L_P(a,b)$ and the 
continuous distance $L_D(a,b)$.

\begin{theorem}
\label{length}
Let $D$ be a compact spacetime domain with the metric normalized so that $\Vol(D)=1$. 
Let $a,b\in D$ be such that $a\prec b$.
Let $P$ denote a discrete spacetime associated with $D$, then, for any $\varepsilon >0$, w.h.p.\ 
\begin{equation}
\label{geodesiclength} 
\vert \frac{L_P(a,b)-c_dL_D(a,b)n^{1/d}}{n^{1/d}}\vert <\varepsilon
\end{equation}
where $c_d$ is a constant which depends only on the dimension. Moreover, given any Riemannian metric which is compatible with the differential structure of $M$, and any $\varepsilon >0$ the maximal chain in $P$ between $a$ and $b$ will be contained w.h.p.\ in an $\varepsilon$-neighborhood of a maximal curve in $D$ between $a$ and $b$.
\end{theorem}

The result has a somewhat tortured history.
It is implicit in \cite{Myrheim}. It was proved in Minkowski space in \cite{BB}, stated explicitly in \cite{BG} 
and proven in dimension $d+1=2$ in \cite{DZ}.
The proof of \cite{DZ}, which is in a completely different context (no mention of spacetime geometry), generalizes 
to arbitrary dimension. Meanwhile, in the physics literature on discrete gravity it seems to retain the status of a conjecture,
\cite{Valdivia, ITR}. A numerical study can be found in \cite{ITR}. 

The constant $c_d$ which appears in the theorem is known only for $d=1$, where according to results of Kerov and Vershik and
of Logan and Shepp, \cite{VK,LS} we have $c_1=\sqrt{2}$. For $d>1$ there are only bounds for $c_d$, \cite{BB}, which are obtained via first moment methods.

The proof of theorem~\ref{length} is fairly elementary. It is first proved for an interval in Minkowski space. In this case
it follows from a simple scaling argument and the transitivity of
the Lorentz transformation group (the isometries) on timelike lines.
The move from the local result to the global one is achieved using finite coverings of $D$ by  
small patches where the metric is nearly constant, and by using the upper semi-continuity of the length functional, \cite{Pe}.

Given the law of large numbers $L_P(a,b)\approx c_DL_D(a,b)n^{1/(d+1)}$, it is natural to ask for a central limit theorem. 
Therefore, we define the random variable error term 
\begin{equation}
\label{deltadef}
\Delta_{D,n}(a,b)=L_P(a,b)-c_DL_D(a,b)n^{1/(d+1)}
\end{equation}
A general concentration result of Talagrand, \cite{Ta},
shows that w.h.p.\ $\Var(\Delta_{D,n})=O(n^{1/(d+1)})$. For $d>1$, essentially nothing is proven, beyond this fact.
One expects that when the maximal curve is unique and entirely 
contained in the interior of $D$, the order of magnitude of $\Delta_{D,n}$ will only depend on the dimension of $D$. One may further expect that
as $n$ goes to infinity, $\Delta_{D,n}$ 
will have the limiting form 
\begin{equation}
\label{beta}
n^{\beta_d/(d+1)}\chi_D
\end{equation} 
where $\beta_d$ is a dimension dependent constant and $\chi_D$ is 
some distribution. When  
$d=1$, and $D=[a,b]$ is an interval in the Minkowski plane, the famous Baik-Deift-Johansson theorem, \cite{BDJ} states that $\Delta_{D,n}$ does have a limiting form as above with $\beta =1/3$ and
$\chi_D=F_2$ where $F_2$ is the Tracy-Widom distribution of fluctuations of the largest eigenvalue in the GUE ensemble.
Results of Baik and Rains \cite{BR1,BR2} show that $\chi_D$ for the triangular domain $a,e,b$ in figure~\ref{fig:rhombus} is $F_4$, the Tracy-Widom distribution of GSE, and the distribution corresponding to the triangle, $a,e,f$
is $F_1$, the Tracy-Widom distribution for GOE, see \cite{TW} for more on these distributions. These results are much deeper than the other results presented in this survey. 

\begin{figure}[tb]
\begin{center}
\vspace{.375in} 
\includegraphics[width=4.5in]{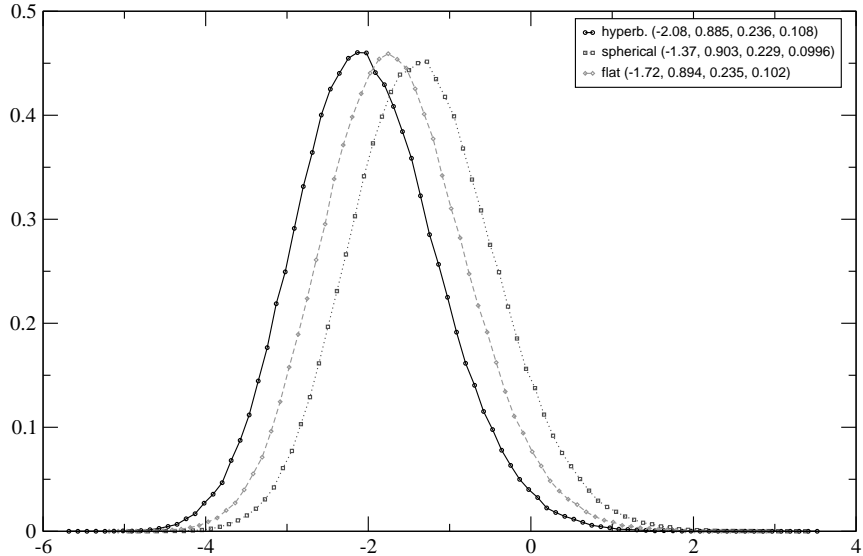}
\end{center}
\caption{$\Delta_n$ for intervals in flat, spherical and hyperbolic space.}
\label{fig:TW} 
\end{figure}

Figure~\ref{fig:TW} shows the distribution of $\Delta_{D,n}/n^{1/6}$ in discrete spacetimes with $500,000$ points associated with a hyperbolic, flat and spherical interval, which are represented by the domain $\vert x\vert +\vert y\vert \leq \pi /8$. Each distribution was composed of roughly $350,000$ samples. For the 
domain in question one can calculate $c_dL_D$ analytically to obtain 
$c_dL_D\approx 1.973\sqrt{n}$ for the hyperbolic metric, $c_dL_D\approx 2.026\sqrt{n}$ for the spherical and $c_dL_D=2$ for the flat case. 
The figure also displays the
average, variance, skew and kurtosis, for the distributions. The respective values for $F_2$ (flat space asymptotics) are approximately 
$-1.771, 0.902, 0.224, 0.093$, \cite{TW}. As can be seen, the numerics suggest that the following hypothesis:

In hyperbolic and spherical spacetime
the distribution $\chi_D$ has the form $\alpha F_2+\gamma $, where $\gamma $ has the same sign as the curvature and $\alpha =1$ if the interval domain is calibrated so that $c_dL_D= 2$. A relation between the shift and curvature is expected since the curvature controls the length of curves which are close to the geodesic.   

Moving to $d>1$ nothing is proven even for the flat interval case. 
The behavior of $\Delta_{D,n}$ has been studied in the context of statistical physics because one of the 
simplest surface growth models, 
known as {\it Polynuclear growth} (PNG) can be described in terms of a discrete spacetime \cite{PS, Pr}. 
In the PNG model one considers a crystal in contact with a super-saturated vapor. Once in a while a nucleus is formed on the existing surface and one assumes that the nucleus spreads evenly in all directions at a constant speed, forming a round 
new layer. When two such layers collide they coalesce. In terms of the PNG model, theorem~\ref{length}
describes the macroscopic shape of the surface while the properties of $\Delta_{D,n}$ describe finer aspects such as roughness 
of the surface. PNG is a model within a larger class of Kardar-Parisi-Zhang growth processes which are expected to exhibit some universal features, including the values of the exponents $\beta_d$. 
Some heuristic theoretical
considerations in physics, \cite{La}
, suggest that $\beta_2 $ and $\beta_3$ have the form $x/(2-x)$ for some $x$ of the form $2/(k+2)$, where $k$ is an odd integer.
Numerical measurements predict
that $\beta_2=0.24$, $\beta_3=0.165$, \cite{Pr}. The closest values predicted by the heuristic argument would be $\beta_2=1/4$
($k=3$) and $\beta_3=1/6$ ($k=5$), however, it is claimed in \cite{Pr} that the numerical value $\beta_2=0.24$ seems significant enough to exclude $\beta_2=1/4$. 
A very interesting prediction of the physical heuristic arguments is that $\Var(\Delta_{D,n})<n^{\varepsilon}$ 
for all $\varepsilon >0$ and $n$ large enough, when $d\geq 4$ (essentially $\beta_d=0$ for $d\geq 4$.). The existence of such a 
critical dimension is subject to great debate in the physics literature, \cite{CM,La,MPPR},
and is currently not supported by numerical evidence. Numerics lead to an estimated value of $\beta_4$ in the range $0.1-0.15$. 
We note that many of the numerics are not computed on discrete spacetime, but rather, on closely related partially ordered sets
in which the light cone is replaced by a ``simplicial light cone''. The main example is the unit $m$ dimensional cube with the
partial order relation $(x_0,\ldots ,x_d)\prec (x_0',\ldots ,x_d')$ iff $x_i\leq x_i'$ for all $0\leq i\leq d$. It is expected that, after proper normalization, the fluctuations of the longest chain for this partial order are the same as those of the flat interval spacetime
(universality). 
Numerical calculations are easier and faster for the simplicial partial order.

Another result which is strongly related to theorem~\ref{length} concerns the behavior of a random chain of length $n$,
as $n$ becomes large. We will state a semi discrete version. 
\begin{theorem}
\label{randomchain}
Let $a,b\in D$
Let $P=\left\{ x_1,\ldots x_n\right\} $ be a discrete 
spacetime associated with $D$ which is conditioned to be a chain between $a$ and $b$ , i.e., $a\prec x_1\prec x_2\prec \ldots
\prec x_n\prec b$, 
then for any $\varepsilon >0$, w.h.p., $P$ is contained in an $\varepsilon $ neighborhood of a maximal curve between
$a$ and $b$.
\end{theorem}
Locally, where maximal curves and timelike geodesics coincide we can think of theorem~\ref{randomchain} as providing a nice 
probabilistic interpretation for Einstein's law of motion, namely, a completely random trajectory of a particle between two endpoints
is likely w.h.p.\ to closely follow a geodesic. So, perhaps God does play dice but as stated by a wise old man, 
``everything is foreseen, yet freedom of choice is given'', Rabbi Akiva, Avot 3:19.

Theorem~\ref{randomchain} was proven in \cite{DZ}, again in a very different context, for two dimensional domains, 
the proof extends to arbitrary dimension. A probabilistic interpretation along these lines to Einstein's equation of motion was 
sought in \cite{DHS}. For similar and related results in the context of relativistic Markov processes and 
Brownian motion see \cite{Du1, Du2, Du3}.

So far we have considered maximal chains in a discrete spacetime. We can consider the analogues problems for anti-chains.
An {\it anti-chain} is a set of points no pair of which is comparable w.r.t.\ $\prec$. In a continuous spacetime this is known 
as an {\it achronal set}. In the same way that chains are related to the concept of timelike curves, 
anti-chains are closely related to the concept of a spacelike hypersurface.
A differentiable hypersurface is said to be spacelike if all its tangent vectors are spacelike. We have noted that geodesics locally
maximize proper time. Proper time may be identified with the induced volume of a timelike curve. Similarly we may consider 
{\it maximal hypersurfaces} which are spacelike hypersurfaces which locally maximize the induced volume among spacelike hypersurfaces.
The corresponding Euler-Lagrange equation which is satisfied by a maximal hypersurface states that the scalar curvature 
vanishes along the hypersurface.
In analogy with theorems~\ref{length} we can state the following result.
\begin{theorem}
\label{antichain}
Given $P$ tobe a discrete spacetime associated with a domain $D$ of unit volume, let $a(P)$ denote the size of a maximal anti-chain in $P$.
Let $a(D)$ denote the maximal volume of an achronal spacelike hypersurface in $D$, then
there exists a constant $c'_d$ such that for all $\varepsilon >0$, w.h.p., 
\begin{equation}
\label{maxanti}
\vert \frac{a(P)-c'_da(D)n^{d/(d+1)}}{n^{d/(d+1)}}\vert <\varepsilon
\end{equation}
Moreover, if $A\subset P$ is an anti-chain of size $a(P)$, then, w.h.p., there is a spacelike hypersurface $N$ of volume $a(D)$
such that $A$ is in an $\varepsilon $ neighborhood of $N$.    
\end{theorem}
Similarly, we have the analog of theorem~\ref{randomchain}, which states that a large random anti-chain in $D$ will cluster around an achronal spacelike hypersurface of maximal volume. 

Geodesics and maximal hypersurfaces are examples of critical points of the volume functional. The Nambu-Goto action in string theory is also given by the volume functional, this time applied to surfaces which have a timelike tangent and a spacelike tangent at each point. Such surfaces describe the (classical) trajectory of a string. As before, one is interested in the critical points of this action. 
It would be interesting to provide some probabilistic interpretation to the critical points via causality preserving maps
from a family of discrete spacetimes. 

We can use theorems~\ref{length} or~\ref{antichain} to provide a very simple procedure for reconstructing the dimension 
of $D$ from a large associated 
discrete spacetime. Several solutions for this problem were proposed, \cite{M,Valdivia,R}, 
however, all of them rely on specific properties of intervals in Minkowski space and do not work 
for general domains in general spacetimes. 

Let $P$ be a 
given discrete spacetime associated with $D$. Randomly subdivide $P$ into two disjoint sets $P_1$ and $P_2$ of size 
$2n/3$ and $n/3$ respectively. We compute the sizes $k_1,k_2$ of longest chains in $P_1,P_2$ and the sizes $l_1,l_2$ of the maximal anti-chains. By theorem~\ref{length} the ratio 
$k_1/k_2$ will approach $2^{1/(d+1)}$ w.h.p., therefore we can use $1/\log_2(k_1/k_2)$ as 
our estimate of the dimension of $D$. Similarly, by theorem~\ref{antichain} the ratio $l_1/l_2$ will approach $2^{d/(d+1)}$ and so $1/(1-1/\log_2(l_1/l_2))$ can serve as a dimension estimator. The basic idea of dividing $P$ into two sets of uneven size 
is rather flexible and can be used in conjunction with many other properties of the partially ordered set in a similar way.
The accuracy of the method depends on the properties of error terms like $\Delta_{D,n}$ or $a(P)-c'_da(D)n^{d/(d+1)}$, 
to the best of our knowledge the latter has not been explored even numerically.
    
Having considered chains and anti-chains we may also consider a well known ``bottom to top'' peeling process which involves both. Let $P$ be a finite 
partially ordered set. We denote by $A_1$ the set of minimal elements of $P$, by $A_2$ the set of minimal elements of $P-A_1$ 
and more generally, by $A_i$ the set of minimal elements of $P-\cup A_j$, $j<i$. The number of such sets is well known to be equal to the size of the longest chain, hence we can analyze its asymptotic behavior using theorem~\ref{length}. The peeling process will figure prominently in the applications which will be described below.

We can symmetrically define the ``top to bottom'' peeling process with $A_i'$ defined recursively 
as the set of maximal elements. We call the set $A_1'$, consisting of the maximal elements in $P$, the {\it maximal layer}. The maximal layer plays an important role in computational geometry, especially in the two dimensional case. Consequently,
the asymptotics of the maximal layer have been studied in depth in the case $d=1$, \cite{Dev,BHLT}. We can consider $D_{max}$,
the set of maximal points of the domain $D$. Generically, $D_{max}$ will be a spacelike curve. We can consider its length $L(D_{max})$ w.r.t.\ the metric $-g$ for which 
it is timelike. We can restate the main result of \cite{Dev} as saying that if $vol(D)=1$, 
w.h.p.\ the size of the maximal layer will be approximately $\sqrt{\frac{\pi }{2}}L(D_{max})\sqrt{n}$, if $L(D_{max})>0$. This result generalizes to any dimension if we replace $L(D_{max})$by $\Vol(D_{max})$ and $\sqrt{\pi /2}$ by a constant, which is currently unknown. 
It is shown in \cite{BHLT} that the error term has a normal distribution with order of magnitude $n^{1/4}$. The case where $\Vol(L_{max})=0$, as with intervals, has been studied 
in the context of the simplicial light cone partial order on the cube, \cite{BDHT,Bar,BT}. 
The expected size of the maximal layer in this case is $\frac{1}{(m-1)!}\ln(n)^{m-1}$,
and the fluctuations are again normal. We can expect similar results to hold for discrete spacetime, with different constants. The size of the maximal layer in an
interval is of interest since it determines the number of edges in the graph representation of a discrete spacetime. 
 
Paper \cite{Dev} also presents some examples of finite volume, non-compact domains, for which the above asymptotic 
estimate does not hold. From the point of view of spacetime geometry, the problem is that in the counter examples
the length function w.r.t.\ $-g$ is not upper semi-continuous at $D_{max}$, a problem which does not occur when $D$ is compact. 
 
For additional reconstruction results the reader may consult \cite{MRS}.
\section{A list of problems}
We present a few problems from different fields of applications.

\

\noindent 1) {\bf Scheduling of I/O (input/output) requests in disk drives: }
Modern disk drives such as the one in a PC or laptop can 
handle multiple requests to read and write data. They can 
rearrange the order of requests so as to handle them 
more efficiently. In practice this is a very important feature 
and in some cases can double the performance of the disk drive. 
The problem of finding the optimal order of the requests is known 
as the {\it disk scheduling} problem. We may also ask for 
the amount of time that will be required to handle 
all the requests. This is known as {\it service time}. Since disk drives rotate at a constant speed we may ask equivalently for the number of disk rotations which are required to handle all the requests. We refer the reader to 
\cite{ABZ, Seltzer,Wilkes} for more information on disk scheduling. 

As a toy model for this problem we may consider the simplistic case in which we assume that 
the read/write head of the disk accelerates and decelerates instantaneously. 
Under these assumptions Andrews, Bender and Zhang (ABZ) gave an algorithm which provides an optimal ordering of requests, 
\cite{ABZ}. 

What is the expected service time of the ABZ algorithm for $n$ requests?

\

\noindent 2) {\bf Connecting-the-dots via Lifschitz functions: } 
The following problem was introduced in \cite{ADX} in the context of estimating parameters for a multi-scale pattern recognition
algorithm. 

Given a 
set of $n$ points sampled via some density in a bounded 
domain in $d+1$ dimensional Euclidean space. 

What is the maximal number of points which can lie on the 
graph of a Lifschitz function (on the first $d$ coordinates) with 
Lifschitz constant $1$?

\

\noindent 3) {\bf Airplane boarding: } We consider passengers boarding an airplane. The input parameters for the process are

\begin{itemize}

\item $u$ -- the average amount of aisle length occupied by a passenger;

\item $w$ -- the distance between successive rows;

\item $b$ -- the number of passengers per row;

\item $D$ -- the amount of time (delay) it takes passengers to clear the aisle,
once they have arrived at their designated row;

\item An airline boarding policy. For example, the policy may first allow passengers from 
the back third of the airplane to board, followed by the front third
and finally the middle third. Another choice which is very popular is to board passengers from the back to the front,
say, the back third, followed by the middle third and finally the front third. 

\end{itemize}

Given $u,w,b,D$ what is a good boarding policy for an airline to pursue?

\

\section{Problem solutions}
We show how the problems presented above are linked to discrete spacetime.

\

\noindent 1) {\bf Scheduling of I/O (input/output) requests in disk drives: } Disk locations can be represented in polar coordinates
$(r,\theta )$, say with $1\leq r\leq 2$ and $0\leq \theta \leq 2\pi$ and $(r,0)$ identified with $(r,2\pi )$. 
We are also given a distribution $p(r,\theta )$ which reflects the popularity of different data in the disk.
We can then assume that the data requests are sampled w.r.t.\ $p$.
We assume for simplicity that the maximal radial speed of the disk is $1$. We may then consider 
the Lorentzian metric 
\begin{equation}
\label{disk}
ds^2=2p(r,\theta )(dr^2-d\theta ^2)
\end{equation} 
on the disk. 
Spacelike curves w.r.t.\ this metric correspond to possible motions of the disk head in a single rotation. Since the volume form for this metric is proportional
to $p(r,\theta )$ we can think of $n$ data requests as forming a discrete spacetime. 
We may apply the peeling process to this discrete spacetime. The ABZ algorithm has 
a very natural interpretation in terms of the peeling process. 
The i'th layer $A_i$ is precisely the set of requests that will be handled in the i'th rotation of the disk. The service time 
will then be the number of layers in the peeling process which according to the results of the previous section is asymptotic to
$L(C)\sqrt{n}$ where $L(C)$ is the length of the maximal curve in the model. Since the boundary of the model is spacelike, the maximal
curve will be a geodesic in this case. 
In disk drives, files are laid out in consecutive tracks, each track corresponding to a certain fixed radius $r$. 
This type of data arrangement 
naturally leads to a popularity density $p$ which depends only on $r$, namely $p=p(r)$. In this case symmetry considerations 
show that the maximal curve is any curve which fixes the angle $\theta $ and calculating the length we get 
that the service time concentrates around
$\sqrt{2}\int_0^1\sqrt{p(r)}dr\sqrt{n}$. Since we have an infinite number of maximal curves the magnitude of the 
error term $\Delta_{D,n}$ can be somewhat larger than expected. When the distribution $p$ is the uniform distribution, it can be shown using results from \cite{Jo,LM} that $\Delta_{D,n}$ has order of magnitude $n^{1/6}\log(n)^{2/3}$. We expect that the variance will actually be smaller in this case, in analogy with extreme value distributions. For more information, see \cite{Ba3,Ba4}.

We need to say a few words about the (non) realism of this model. The assumption that the disk does not accelerate or decelerate 
is far from true. The model which we have used predicts that a disk can handle an order of magnitude of $\sqrt{n}$ requests
per rotation,
a real disk will not handle more than 2-3 requests per rotation regardless of the number of requests, due to some fixed time delays.
Nonetheless, the type of algorithms that are used in this case (ABZ) are indicative of successful algorithms in
more complex and realistic models, therefore one does gain an understanding of the full problem by considering 
these simple cases.

\ 

2) {\bf Connecting-the-dots via Lifschitz functions: } The relation between this problem and spacetime geometry becomes clear 
once we note that by definition the graph of a Lifschitz function is a spacelike hypersurface in Minkowski space. 
In order to think of the points as points in a discrete spacetime we have to match the volume form of the metric to 
that of the point distribution so we introduce 
the conformally flat metric 
\begin{equation}
ds^2=p(x,t)^{2/(d+1)}(dt^2-\sum_idx_i^2).
\end{equation} 
By theorem~\ref{antichain} the largest number of points
will be picked in an $\varepsilon $ neighborhood of a maximal hypersurface 
in the domain and the number of points will be proportional to the volume of the hypersurface. We also notice that this problem 
is a generalization of the previously discussed disk scheduling problem.   

\

3) {\bf Airplane boarding: } We represent passengers by their position in the boarding queue $x$ and their row number $y$, both normalized to produce a point $(x,y)$ in the unit square. 
The airline boarding policy (and the way passengers react to it) leads to a distribution 
$p(x,y)$ on the joint row/queue position. For example, assume that the policy is
to board the back third of the airplane, followed by the front third
and finally the middle third. The corresponding distribution $p(x,y)$
is given by $p(x,y)=3$ on the sub-squares $[0,1/3]\times [2/3,1]$,
$[1/3,2/3]\times [0,1/3]$, $[2/3,1]\times [1/3,2/3]$, and
$p(x,y)=0$
elsewhere. The back-to-front policy with 3 equal group sizes will lead to a distribution with $p(x,y)=3$ on the subsquares $[0,1/3\times [0,1/3]$, 
$[1/3,2/3]\times [1/3,2/3]$ and $[2/3,1]\times [2/3,1]$.

\begin{figure}[tb]
\centerline{\includegraphics[width=4.5in]{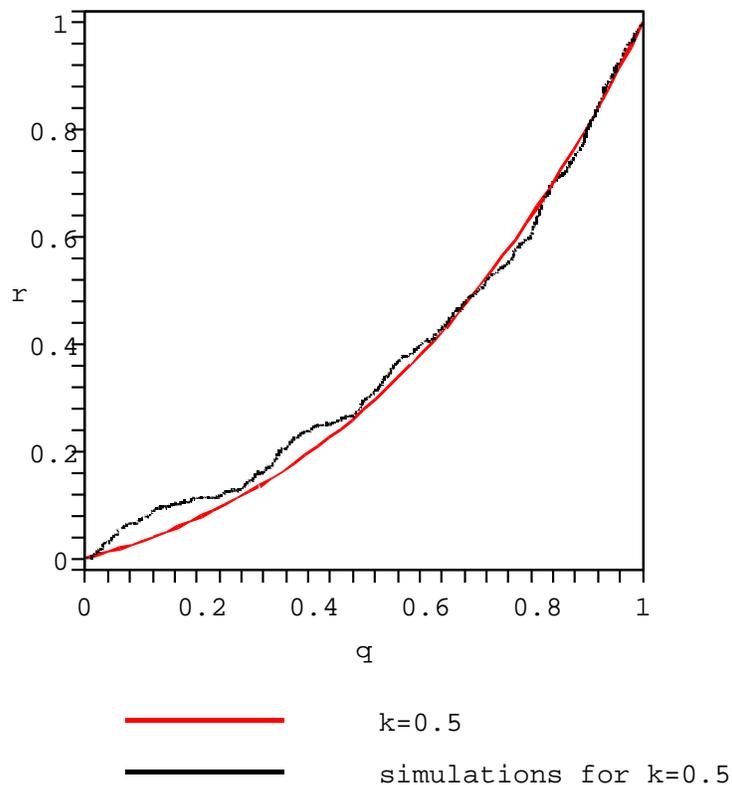}}
\caption{The maximal curve and the longest chain for $10^6$ passengers, when $k=0.5$}
\label{figure3} 
\end{figure}

\begin{figure}[tb]
\centerline{\includegraphics[width=4.5in]{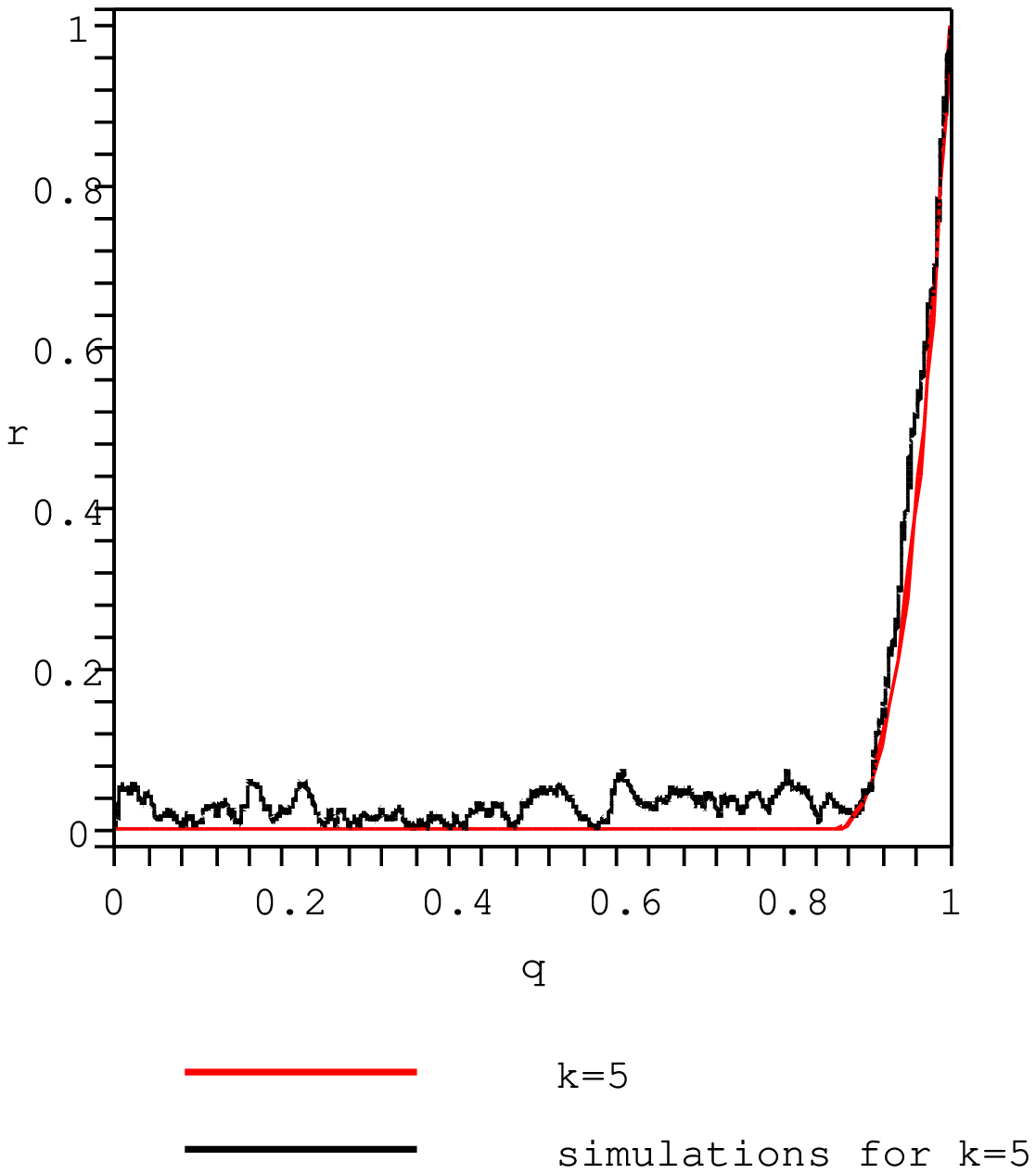}} 
\caption{The maximal curve and the longest chain for $10^6$ passengers, when $k=5$}
\label{figure4} 
\end{figure}

We define $k=\frac{ub}{w}$. $k$ measures the congestion of passengers in the airplane. It is the reciprocal of the 
portion of all passengers which can comfortably stand along the entire length of the airplane aisle. 
In real airplanes, $k$ equals roughly $4$. We also define 
$\alpha (x,y)=\int_y^1p(x,z)dz$. $\alpha$ measures the density of passengers which want to pass by the passenger $(x,y)$
to get to their seats. Given $p$ and $k$ we
define the Lorentzian metric 
\begin{equation}
\label{airplane}
ds^2=4D^2p(x,y)(dx(dx+k\alpha (x,y)\,dy))
\end{equation}
Since $p(x,y)$ is proportional to the volume form, we can think of the passengers as a discrete spacetime sampled from the unit square equipped with this metric. It can be shown, \cite{Ba1} that the causal relation $\prec $ associated with the metric (\ref{airplane}) asymptotically coincides with the notion of passengers blocking
other passengers from getting to their seats. 
Consequently, the boarding process roughly coincides with the bottom-to-top peeling process applied to this discrete 
spacetime. The first layer $A_1$, corresponds to passengers which can sit down without being blocked by other passengers. 
More generally, the set $A_i$ asymptotically corresponds to the passengers which can sit unobstructed once the passengers 
in $\cup A_j$, $j<i$, have sat down. The boarding time is then given by $L(C)\sqrt{n}$, where $C$ is the longest (maximal) 
curve in the unit square with the metric (\ref{airplane}. In figures~\ref{figure3} and~\ref{figure4}, we see a comparison of the maximal curve, with the longest chain
of passengers blocking each other, for a huge airplane with $10^6$ passengers and congestion $k=0.5$ and $k=5$ respectively. 
We must admit that when the number of passengers is $200$ or so
the maximal curve does not follow the geodesic well. Furtheremore,
it turns out that large portions of the maximal curve typically lie on the boundary of the support of $p(x,y)$.   
This causes significantly greater variations in boarding time and makes the error term $\Delta_{D,n}$ much larger, \cite{Ba1}.
Nonetheless, rather miraculously the ranking of boarding policies which is produced by the estimate  $\ell (C)\sqrt{n}$ 
nearly coincides with the rankings given by very detailed computer simulations of the boarding process, \cite{Ba2,VB}. The methods above led to a detailed understanding of back-to-front policies as practiced by many airlines. Such policies are parameterized by the infinite dimensional simplex, consisting of sequences $r_0=0<r_1<\ldots <r_{m-1}<1=r_m$, for any positive integer $m$. Given such a sequence,
$n$ passengers, with $d$ passengers per row, the corresponding policy boards passengers in rows $r_{m-1}n/d$ to $r_mn/d=n/d$,
followed by passengers in rows $r_{m-2}n/d$ to $r_{m-1}n/d$ and so on. 
Using these methods it was shown in \cite{Ba5} that for $k>1$ the expected boarding time of any back-to-front policy
is at most 
\begin{equation}
\label{back-to-front}
\frac{\sqrt{k}+\frac{1-\ln(2)}{\sqrt{k}}}{\sqrt{k-1}}
\end{equation}
faster than random boarding (no airline policy). 
For the realistic value of $k=4$ it shows that we cannot obtain a speed up of more than 20\% in boarding time using back-to-front
and in fact most back-to-front policies yield performance which is worse than random. The estimate~\ref{back-to-front}
which is based on constructing long (but not maximal) curves whose length satisfies a recursive inequality, is not tight.
We are currently exploring more sophisticated recursions which will yield better bounds for any given $k$. We conjecture that for $k=4$ no back-to-front policy will yield a reduction in boarding time of more than 5\% . For popular accounts of the airplane boarding problem see \cite{Nature,Newscientist,Za}.

{\em Acknowledgements}: We thank Vassilii Khachaturov for producing figure 2 and for making many suggestions for improving the presentation.



\end{document}